\documentclass[11pt]{article}
\usepackage{amssymb,amsmath,amsxtra,cite,bm,color,mathrsfs}
\usepackage{graphicx}   
\usepackage{float}
\usepackage[hidelinks]{hyperref}
\usepackage{xcolor}
\hypersetup{
	colorlinks,
	linkcolor={red!50!black},
	citecolor={blue!50!black},
	urlcolor={blue!80!black}
}
\begin{document}
\begin{center}
{\large\textbf{On swimmer's strategies in various currents}
 }
\\[\baselineskip]
{\sffamily Amir~Aghamohammadi$^{a,}$\footnote{e-mail: mohamadi@alzahra.ac.ir} ,
	Cina~Aghamohammadi$^{b,}$\footnote{e-mail: ca6941@princeton.edu},
	Saman~Moghimi-Araghi$^{c,}$\footnote{e-mail: samanimi@sharif.edu}}\\[\baselineskip]
$^{a}$ Department of Fundamental Physics, Faculty of Physics, Alzahra
University, Tehran, Iran,\\
$^{b}$ Princeton Neuroscience Institute, Princeton University, Princeton, NJ 08540,\\
$^{c}$ Physics department, Sharif University of Technology, P.O. Box 11155-9161, Tehran, Iran
\end{center}
\vspace{2\baselineskip}

\begin{abstract}
Finding an optimum strategy to reach a certain destination by swimming in a background flow is an interesting question which leads to non-trivial results and swimming paths. Here we consider different strategies for various types of surface currents, including parallel currents, and currents resulting from spring sources, and sinks. Swimmers may instinctively swim toward the destination point. It turns out that this idea may not lead to the desired result in the presence of the background current. We will show in some cases the best strategy is to swim towards a certain point other than the actual destination. A different strategy may be to obtain the path of the least time and then follow the minimized path. We study this alternative strategy, too.    
\end{abstract}
\vspace{2\baselineskip}
\section{Introduction}

 We will discuss different strategies for swimming in a background flow field in this article. Swimming in a pool differs from swimming in open water. This is mainly due to currents. The swimmer would swim straight to his (her) destination point if the water in the lake were still. When there are currents, swimmers may instinctively swim toward the destination point. However, when there are currents, this is not a good idea. 

A geodesic is the shortest possible distance curve on a surface.  
The most common example of a geodesic is a straight line in Euclidean geometry or any arc of the great circles on a sphere.
This can be rewritten as a moving particle on a surface with a given speed, and ask about the shortest path or minimum time required to reach the destination. One may extend the problem by adding a background flow field. This extended problem was studied by Zermelo \cite{Zermelo}.
He obtained a  partial differential equation, now known as Zermelo's equation \cite{Bryson}. 
The same approach is also used for the geometry of sound rays in a wind \cite{gibbons}. 
Although there is no general solution for Zermelo's equation, and finding an exact solution is nearly impossible in the vast majority of cases,  the problem may be solved numerically.

Problems of this nature can also be found in other contexts. The movement of animals in the air and water may be affected by experienced flow is a similar problem. As an example see \cite{McLaren}. Control theory was used to develop a benchmark for time-minimizing (optimal) orientation there.

Another interesting analogous problem that at first glance may seem irrelevant is the pursuit problem. There, we have two objects: a pursued and a pursuer. Pursued moves along a trajectory while the pursuer follows. See e.g. \cite{Nahin,Bernhart,Richardson,Marshall,Mungan}. Usually, the aim of these problems is to determine the trajectory of the pursuer. Pursuers usually move toward pursuits instinctively.
In \cite{Shunyakov}, it is shown that curvilinear motion on an inclined plane \cite{Shunyakov,CA} is analogous to pursuit. 
Imagine a swimmer moving toward a fixed destination in the presence of a current. The swimmer's speed in still water is $w$, while the surface layer is flowing at a constant velocity of ${\bm u}$.  In a reference frame  moving with the current, this problem transforms into a standard chase problem involving a pursuer chasing a  pursued running with a constant velocity. 
In view of the probable  analogies between each of these problems, analyzing any of them may help us gain a deeper understanding. 

The ability to navigate in an unsteady flow field is crucial to a wide range of robotic applications. In \cite{Gunnarson}, deep reinforcement learning was used to find time-efficient paths through an unsteady, two-dimensional flow field. It is seen that simpler strategies such as swimming toward the target largely fail at the task.

There are currents in open water, including those in rivers, oceans, and seas. Some instances include rip currents, which pose a great threat to swimmers. 
A rip current moves directly away from the shore like a river running out to sea.
To escape a rip, it is dangerous for a swimmer to use the wrong strategies, such as fighting the current. In \cite{Withers}, an 'energetic cost' strategy to escape from a rip current is presented.
In this article, different types of strategies for a swimmer who is going to reach a destination are studied. 
We also consider different types of currents such as parallel current, spring source, and sink.
The concept of taking non-trivial strategy to arrive at the desired point in a given background current make it quite interesting and challenging to an undergraduate student.
The problem and examples addressed here may be of interest in the framework of an undergraduate course in mechanics. Additionally, we believe that our calculations provide skills that complete the formal training. 

The organization of this work is as follows. Section 2 is devoted to navigation in the parallel current. There may be different strategies: swimming directly toward the destination,
picking a guide and swimming directly toward the chosen guide, and eventually swimming along the path of least time. The strategy of the path of least time is equivalent to having a flag or guide point at infinity. In sections 3 and 4, we assume there exists a source or a sink that makes radial currents. In Section 5, mentioned strategies are investigated when there exists a source or a sink. And finally, section 6 is devoted to more sources and sinks.

\section{Parallel current}
Imagine that the water in a lake isn't stagnant and that there is a current. This might be the case of a typical river. The surface layer is  flowing at a constant and uniform velocity ${\bm u}$. The swimmer's speed in still water, $w$, may be greater, smaller, or equal to the current speed, $u$.  The case of $w=u$ is very interesting from pedagogical point of view. Therefore, although it is quite a special case, we will begin with it and analyze it more thoroughly than other cases. 

\subsection{$w=u$}\label{3.1}
\begin{figure}[h]
	\centering
	\includegraphics[width=0.6\linewidth]{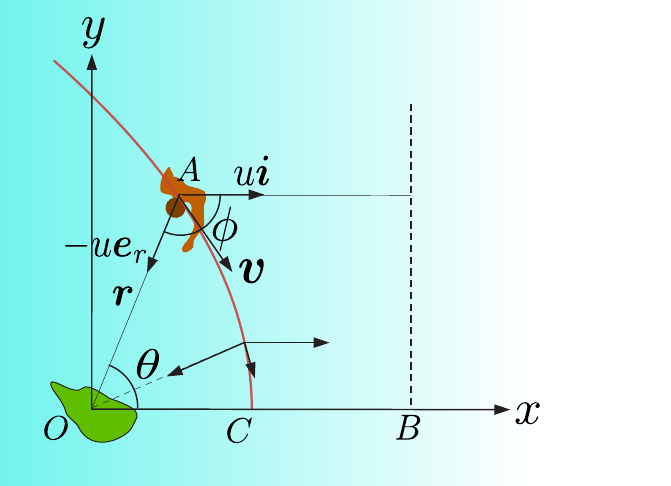}
	\caption{ The swimmer's path 	is a parabola with the  destination as its foci, if he/she swim toward destination.		}
	\label{f1}
\end{figure}

There exist various strategies for the swimmer which we discus later, but  
the instinctive strategy for a swimmer  might be to swim towards point $O$.
Take the $x$ axis in the current direction, and set $O$ as the origin.
Then the swimmer's velocity, ${\bm v}$, is a vector that's made by adding two other vectors,  $-u \dfrac{\bm{r}}{|\bm{r}|}=-u{\bm e}_r$,  the swimmer's velocity with respect to water
which is towards the point $O$, and the other one
$u \bm{i}$ is the current velocity. The angle between these two vectors is $\phi $, and the swimmer's distance to the point $O$ is $r$.
See figure (\ref{f1}). 

It is seen that,
\begin{align}\label{rdot1}
	&\dot r= -u(1+\cos \phi)=-u(1-\cos \theta),\\
	&\dot x= u(1+\cos \phi)=u(1-\cos \theta).
\end{align} 
Therefore, $\dot x+\dot r=0$ or  $x+r=x_0+r_0=:L$ is a constant, which is the general equation of a parabola.  So the path of the swimmer is a parabola that starts from the point $A$, and the nearest point to $O$ along the swimmer's path is $C$. Actually,  in this strategy, the swimmer never passes the point $O$. Let us work out the path of the swimmer explicitly. 
\begin{equation}
	r+r\cos \theta=L\quad \rightarrow \quad r=\frac{L}{1+\cos \theta},
\end{equation}
from which  we can get the swimmer's radial speed
\begin{equation}
	\dot r=\frac{L\, \dot \theta\sin \theta}{(1+\cos \theta)^2},
\end{equation}
which compared to
(\ref{rdot1}),
$\dot \theta$
can  be obtained to be 
\begin{equation}
	\dot \theta=\frac{-u}{L}\sin \theta (1+\cos \theta).
\end{equation}
Now let's see how long it takes for the swimmer to reach the point $C$. In order to get it, we use the relation for $\dot \theta$ and arrive at the following integral
\begin{equation}
	T=\int_0^{T}\mathrm{d}t=\int_{\theta_0}^0\frac{-L \mathrm{d}\theta}{u\sin \theta  (1+\cos \theta)}.
\end{equation} 
This integral can be solved analytically, but there is no real need to do that. 
Let us denote the time to reach $C$ from its nearby points by $T_\epsilon$
\begin{align}
	T_\epsilon=&\int^{0}_\epsilon\frac{-L \mathrm{d}\theta}{u\sin \theta  (1+\cos \theta)}\cr
	\approx &\int^{0}_\epsilon\frac{-L\, \mathrm{d}\theta}{2u \theta  }=\frac{-L}{2u }\ln\theta \big\vert^{0}_\epsilon.
\end{align} 
$T_\epsilon$ is infinite. 
 Considering the fact that $T>T_\epsilon$, T is also infinite.
As the swimmer approaches the x-axis, the size of his (her) speed becomes smaller, and  the swimmer never reaches the point $C$. 

\begin{figure}[h]
	\centering
	\includegraphics[width=0.5\linewidth]{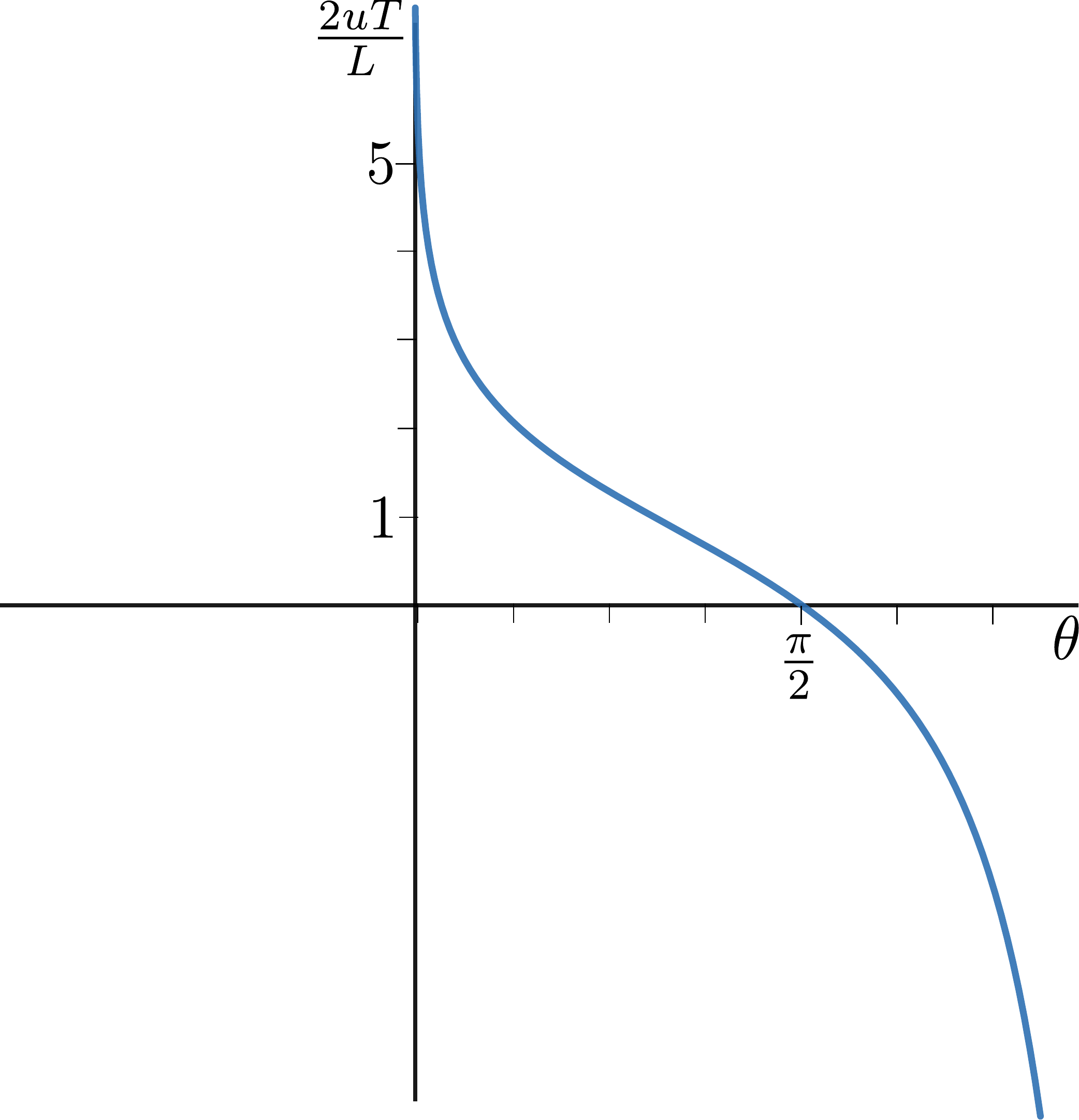}
	\caption{Dimensionless time
		 $\tau(\theta_0):=\dfrac{2uT_{\theta_0,0}}{L}$, versus $\theta_0$.
	}
	\label{f2}
\end{figure}

In order to have a complete discussion, let us compute precisely how long it takes for a swimmer to traverse an arbitrary distance on the path: two point on a parabola.
We need to obtain the following integral
    \begin{align}
	T_{\theta_1,\theta_2}=&\int_{\theta_1}^{\theta_2}\frac{-L \,\mathrm{d}\theta}{u\sin \theta  (1+\cos \theta)}\cr
	=& \dfrac{L}{2u}\,\left(\ln \left|\tan \left(\frac{\theta}{2}\right)\right|+\dfrac{1}{2}\tan ^2\left(\frac{\theta}{2}\right)\right)\Big\vert_{\theta_2}^{\theta_1}.
\end{align} 
Where $\theta_1$ corresponds to the  starting point, and  $\theta_2$  corresponds to the destination.
Here we have used $\sin \theta= 2\sin \theta/2\cos \theta/2$,
and
$1+\cos \theta= 2\cos^2 \theta/2$, and the change of the variable
$g:=\tan \theta/2$. The initial distance $L$, and the speed
$u$, can be used to construct a characteristic time, $\dfrac{L}{u}$, which can be used to de-dimensionalizing  time as  $\tau(\theta_0):=\dfrac{2uT_{\theta_0,0}}{L}$. 
figure (\ref{f2}) depicts dimensionless time $\tau(\theta_0)$ versus $\theta_0$. As can be seen from the figure, the time to reach the $x$ axis  from the point, $\theta_0=\dfrac{\pi}{2}$ and
$r_0=L$, is infinite. So the strategy that the swimmer swims towards the destination is a wrong choice. Later, we will discuss some better strategies, however, we first consider the same strategy when $w\ne u$.

\subsection{$w\ne u$}
\begin{figure}[h]
	\centering
	\includegraphics[width=0.75\linewidth]{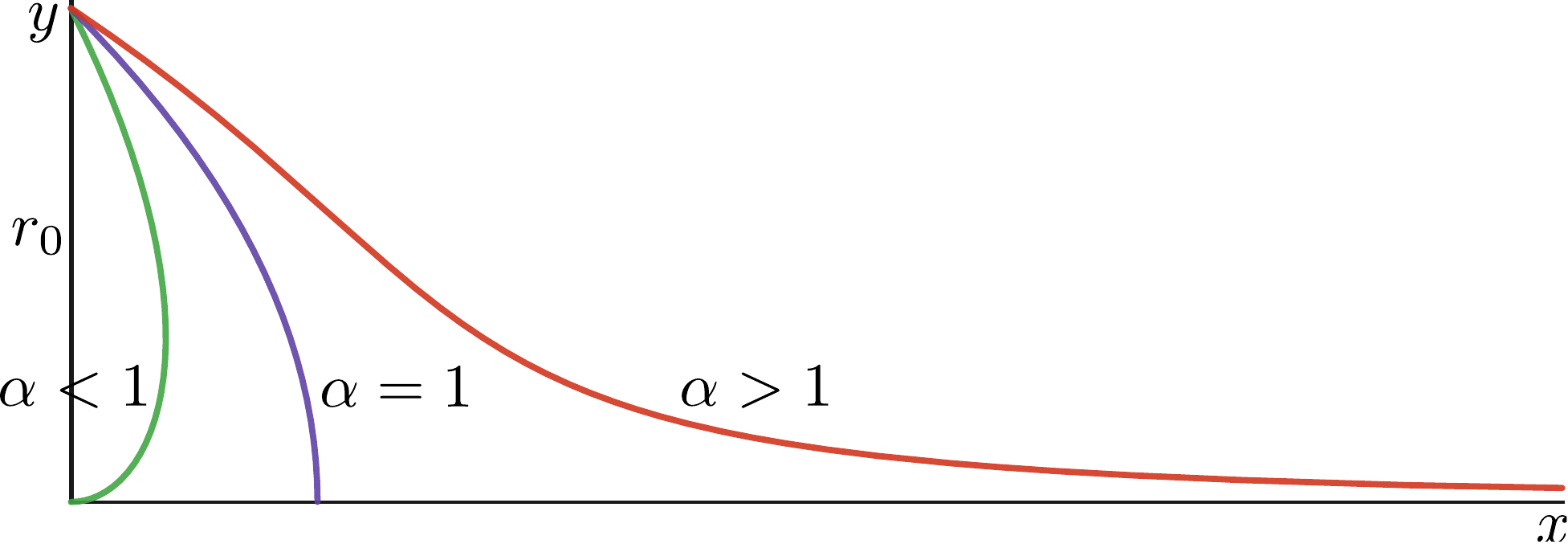}
	\caption{   Path of a swimmer for various  values of 
	$\alpha$ and  according to the initial conditions,
$\theta_0=\dfrac{\pi}{2}$ and $r(0)=r_0$. 
	}
	\label{f4}
\end{figure}

Now suppose the speed of the swimmer is not equal to 
$u$. Then total velocity is 
\begin{equation}
	{\bm v}=-w{\bm e}_r+u \bm{i},
\end{equation}
which leads to the following velocity components in polar  and Cartesian coordinates
\begin{align}\label{polv01}
	\begin{cases}
		\dot r= -w+u\cos \theta\\
		r\dot \theta=- u\sin \theta.
	\end{cases}&\qquad 	
	\begin{cases}
		\dot x=  u-w\cos \theta\\
		\dot y=- w\sin \theta.
	\end{cases}
\end{align}
Dividing the velocity components in polar  coordinates and defining the dimensionless quantity $\alpha:=\dfrac{u}{w}$, we arrive at 
\begin{align}\label{polv02}
	\dfrac{1}{r}\,\dfrac{{\rm d}r}{{\rm d}\theta}=&\dfrac{(1-\alpha \cos \theta)}{\alpha \sin \theta}\cr
	=&-\cot \theta +\dfrac{1}{\alpha \sin \theta}.
\end{align}
This equation can be solved:
\begin{align}\label{polv03}
	\ln\left( \dfrac{r}{r_0}\right)=&\dfrac{1}{\alpha}\,\ln\big\vert\dfrac{\tan(\theta/2)}{\tan(\theta_0/2)}\big\vert
	-\ln\big\vert\dfrac{\sin\theta}{\sin \theta_0}\big\vert,\cr
	r=&r_0\, \dfrac{\sin\theta_0}{\sin \theta}\, \left( \dfrac{\tan\theta/2}{\tan \theta_0/2}\right)^{1/\alpha} 
\end{align}
In figure (\ref{f4}), the path of swimmer for various  values of 
$\alpha$ and  according to the initial conditions
$\theta_0=\dfrac{\pi}{2}$ has been drawn. If
$w<u$ ($\alpha>1$), the swimmer can not even reach to $x$ the axis, and at large times, it can only swim against the direction of $\bm u$ moves asymptotic to $x$ axis and 
away from the destination. Only when 
$w>u$ ($\alpha<1$) the swimmer reaches its destination. 
\\
Swimming time can be calculated exactly. Using eq. 
(\ref{polv01}), one arrives at 
\begin{equation}
	w\dot r+u\dot x=u^2-w^2.
\end{equation}
Integrating the above equation one arrives at
\begin{align}	
	wr+ux=wr_0+ux_0+(u^2-w^2)t,\cr
	r=\dfrac{r_0(1+\alpha \cos \theta_0)+(\alpha^2-1)wt}{(1+\alpha \cos \theta)}.
\end{align} 
As it is seen from (\ref{polv01}), $\dot r<0$ for $\alpha< 1$. Thus  $r$ is a decreasing function of time,  for $\alpha< 1$. As $r$ tends to zero,  time approaches $T$
\begin{equation}
	T=\dfrac{r_0(1+\alpha \cos \theta_0)}{w(1-\alpha^2)}. 
\end{equation}
\subsection{Second strategy: Guide}
In the previous section we showed that the simple strategy to swim towards the destination works only if  the swimmer's speed $w$ is greater than $u$. Let us assume that $w$ is equal to $u$ and propose another strategy that may lead to arrival at the destination. We state the new strategy as follows:  Instead of swimming towards the actual destination, swimmer may pick a guide point and swim towards it. In other words, we put a fictitious flag at some certain point and the swimmer always try to reach the flag. From what we have obtained in previous sections, it is clear that the starting point, $A$, and the destination, $O$, are located on a parabola, whose  focus is the guide point.

\begin{figure}[h]
	\centering
	\includegraphics[width=0.7\linewidth]{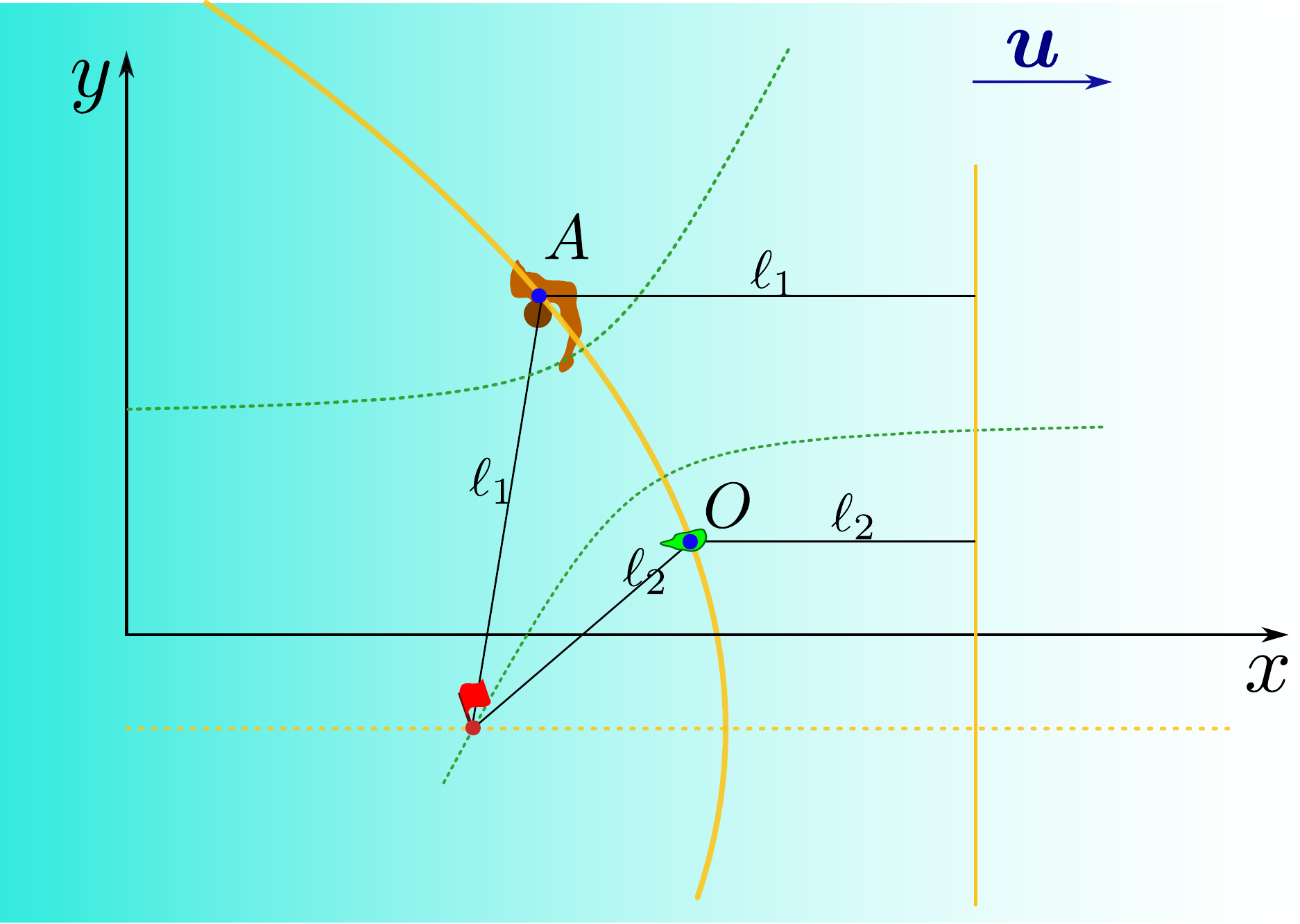}
	\caption{ The path of the swimmer will be a parabola with the flag at its foci if he/she  swims toward the guide, shown by a red flag.   The loci of the points we should set the flag is  a hyperbola (shown by green dashed line) for which two points, initial and destination ($A$ And $O$)  are its foci. 
	}
	\label{f3}
\end{figure}

With that problem in mind, the issue now becomes a completely different problem: How many parabolas are there which passes  the points
$A$ and $O$? And where are the loci of their foci (the guide's position)? In figure (\ref{f3}) a parabola and its focus, which is shown by a flag, is plotted. 
The axes of symmetry for any of these parabolas run parallel to the $x$ axis and  all of their directrices are parallel to 
the $y$ axis.  
There are infinite number of parabolas passing from the points
$A$ and $O$.
Let us show the distance of the point $A$ ($O$) from one of the directrices by $\ell_1$ ($\ell_2$). The distance of these points from the corresponding foci, shown by a flag in figure (\ref{f3}), will be of the same values. If $x_O-x_A=a$, then 
\begin{equation}
	\ell_1-\ell_2=a.
\end{equation} 
Thus the loci of  the foci of these parabolas are  hyperbolas for which two points of $A$ and $O$ are their foci. 

In summary, if the swimmer take any point of this hyperbola as the guide  and sets there, say, a flag, he/she will reach the destination. Asymptotically the swimmer may take a point on the hyperbola that is very far. This simply means that the direction of his (her) velocity remains fixed and the path is a straight line.
\subsection{Third strategy: Time minimization}
\begin{figure}[h]
	\centering
	\includegraphics[width=0.5\linewidth]{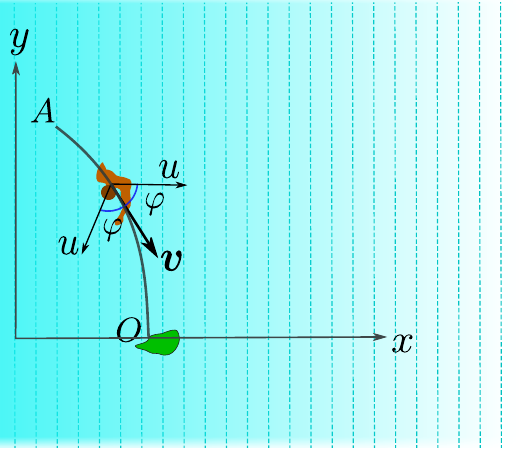}
	\caption{  Calculus of variations can give us the path that minimizes time needed to  reach the destination. 		}
	\label{f5}
\end{figure}

\begin{figure}[h]
	\centering
	\includegraphics[width=0.7\linewidth]{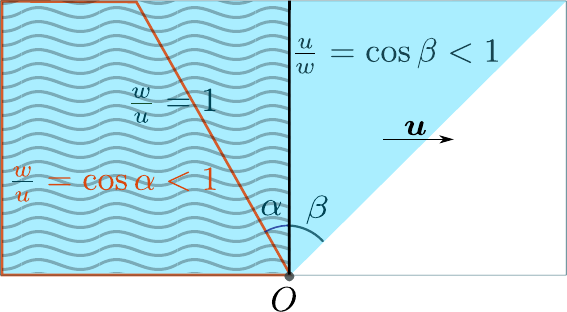}
	\caption{ If $w>u$, from anywhere in the blue trapezoid, the swimmer can reach the point $O$. If both speeds are equal, $w=u$, then the wavy region shows all the initial points where the swimmer can reach the point $O$. And finally, if $w<u$, the swimmer can reach $O$, only when initially it is in the trapezoid with the orange boarder.
	}
	\label{f6}
\end{figure}

Third Strategy is based on finding the path that minimize the time needed to  reach the destination. In order to find the desired path, we may  use the techniques of calculus of variations. Let us first  assume $w=u$. Then the swimmer's speed is $v=2u \cos \varphi$, where 
\begin{equation}
	y'=\tan(\pi-\varphi), 
\end{equation}
is the slope of the path, denoted by $y(x)$.   See figure (\ref{f5}).
The time of swimming is 
\begin{align}
	T=&\int {\rm d}t=\int \frac{{\rm d}s}{v}=\int \frac{{\rm d}s}{2u\cos \varphi}\cr\label{var}
	=&-\dfrac{1}{2u}\int {\rm d}x\, (1+y'^2)=:\int  {\rm d}x\, J(y').
\end{align}
Here we have used of 
\begin{align}
	{\rm d}s=&({\rm d}x^2+{\rm d}y^2)={\rm d}x\sqrt{1+y'^2}\\
	\cos \varphi=&\frac{1}{\sqrt{1+\tan^2\varphi}}=\frac{1}{\sqrt{1+y'^2}}.
\end{align}
Then Euler's equation
\begin{align}
	\dfrac{\partial }{\partial x}\left( \dfrac{\partial J}{\partial y'}\right)-\dfrac{\partial J}{\partial y} =0,
\end{align}
results in $y'$ being constant. Hence, the curve that minimizes the time is a straight line between two points $A$ and $O$.
A similar argument can be made for $w\neq u$. 
In the same manner as equation (\ref{var}), $J$ only depends on $y'$, resulting in a straight line path.  Note that even the path of least time strategy is equivalent to having a flag or guide point at infinity.
If there is an answer to such a problem, then it is a straight line.
However, in some cases, the swimmer may not be able to reach the desired point.
See figure (\ref{f6}).
It is always possible for the swimmer to reach point $O$ from any point inside the blue trapezoid, shown in figure (\ref{f6}), provided $w>u$. 
If $w=u$, from  anywhere in the wavy region can 
reach $O$. And finally, if $w<u$, the swimmer can reach $O$ only if it is initially in the trapezoid with orange boarder.

\section{Source}
\begin{figure}[htpb]
	\centering
	\includegraphics[width=0.48\linewidth]{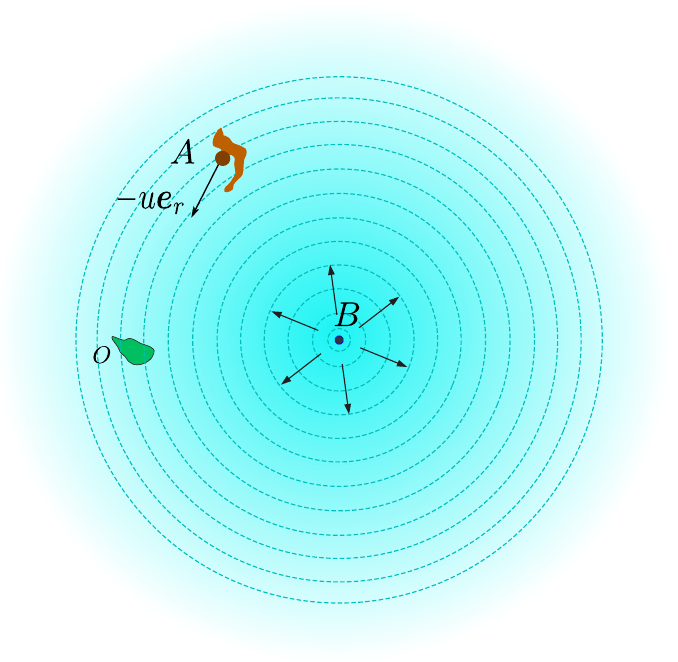}
	\includegraphics[width=0.48\linewidth]{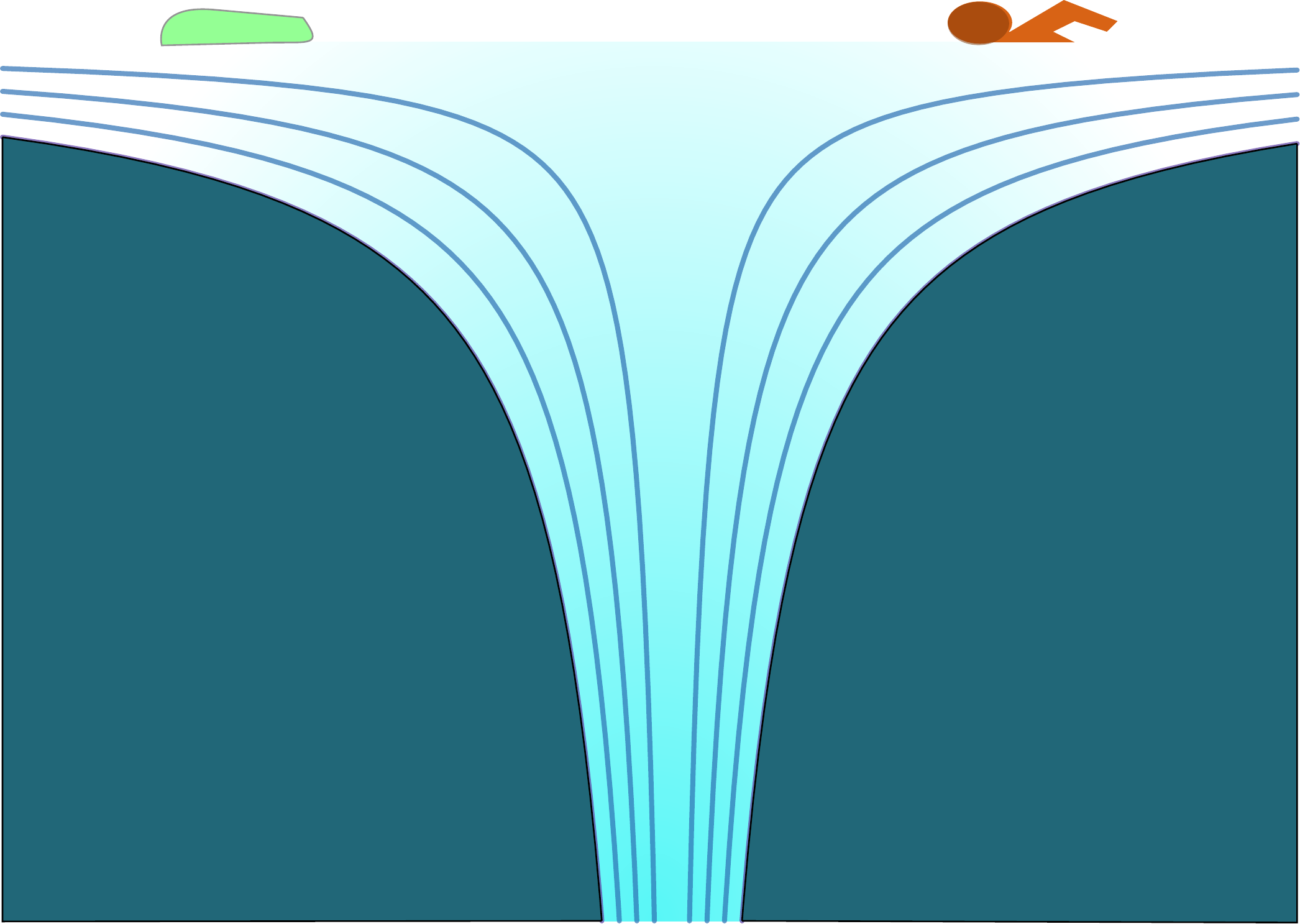}
	\caption{  	 Swimming toward a destination, $O$, in the presence of a source at $B$.  
		}
	\label{f7}
\end{figure}

Up to now, we have considered flows with uniform velocity. We would like to investigate some other types of surface currents that a swimmer may encounter. Another kind of velocity field which turns out to be interesting, at least from the pedagogical point of view, is a `source current', by which we mean the surface velocity of water has a constant magnitude, however, it radially flows outward from a specified point, the source point.
Suppose there's a spring at the bottom of a lake that makes the water on the surface of the lake move away from the source point $B$ with a radial velocity of $u$. The assumption of uniform magnitude of radial velocity is not physical for a uniform depth lake. The condition of incompressibility of water is incompatible with such an assumption. This velocity field can be created by having a spring, but at the same time the lake bottom is not flat. The lake's depth should decrease as $r^{-1}$ from the point that the water leaves the spring, where $r$ is the distance from the spring.
See figure (\ref{f7}).

\begin{figure}[htpb]
	\centering
	\includegraphics[width=0.65\linewidth]{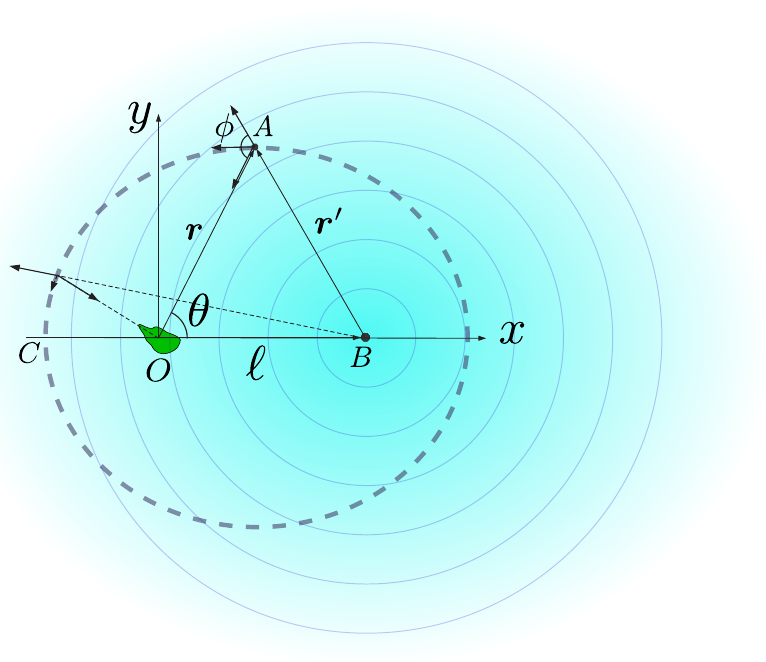}
	\caption{   The swimmer's path 	is an ellipse with the source and destination as its foci ($B$ and $O$), if he/she swims toward destination.	
	}
	\label{f8}
\end{figure}

Let us assume that the velocity of the swimmer, $w$, is equal to $u$. The swimmer's velocity  has two parts, $-u \dfrac{\bm{r}}{|\bm{r}|}=-u {\bm e}_r$ 
and $u \dfrac{\bm{r}'}{|\bm{r}'|}=u {\bm e}_{r'}$, where ${\bm e}_r$ ( ${\bm e}_{r'}$)
is the radial unit vector with respect to the point $O$ ($B$). 
See figure (\ref{f8}). 
The velocity components of the swimmer along $OA$ and $BA$ are
\begin{align}
	&\dot r= -u(1+\cos \phi),\\
	&\dot r'= u(1+\cos \phi),
\end{align} 
which gives $\dot r+\dot r'=0$. Swimmer move on a path for which $r+r'$, the sum of its distances from source and destination ($B$ and $O$),  remains constant, e.g. $L$. This is the general equation of an ellipse. The swimmer's path is an ellipse with $O$ and $B$ as its foci. $r'$ can be written as 
\begin{equation}
	r'=\sqrt{r^2+\ell^2-2r\ell\cos \theta}=L-r,
\end{equation} 
from which we obtain
\begin{equation}
	r=\frac{L^2-\ell^2}{2(L-\ell\cos \theta)}=:\frac{E}{1-e\cos \theta},
\end{equation} 
where
\begin{equation}\label{defEe}	
	E:= \frac{L^2-\ell^2}{2L}, \quad e:=\frac{\ell}{L}.
\end{equation} 
Polar components of the swimmer's velocity are
\begin{align}
	&\dot r= -u(1+\cos \phi),\\
	&r \dot \theta= u\sin \phi.
\end{align} 
Using the following relation between the angles $\phi$ and $\theta$ 
\begin{equation}
	\frac{\sin \phi}{\ell}=\frac{\sin \theta}{L-r}
\end{equation} 
we arrive at 
\begin{equation}\label{thetT}
	\dot \theta=\frac{u\ell\sin \theta}{r(L-r)}. 
\end{equation} 
As $\theta$ (or $\phi$) tends to $\pi$, $\dot \theta$, and  $\dot r$ approaches zero. It occurs at the point $C$, where the ellipse intersects the axis $x$, and the swimmer stops.
Although it starts heading in the direction of point $O$, eventually it will stray away from point $O$ until it finally stops at point $C$ a distance of $\dfrac{L-\ell}{2}$ away from point $O$.
Integrating (\ref{thetT}), we obtain the time needed to reach $C$.
\begin{equation}
	T=\int_0^{T}\mathrm{d}t=\int_{\theta_0}^\pi\frac{\mathrm{d}\theta\ r(\theta)(L-r(\theta))}{u\ell\sin \theta}.
\end{equation}
Using similar arguments of subsection (\ref{3.1}), the time it takes for the swimmer to reach that point from the vicinity of the $C$ to it, is infinite, and as a result, the total time to reach the point $C$ is also infinite.

However, the time needed to reach from any point on this path  with $\theta_1$ to another one, $\theta_2$, can be obtained exactly 
\begin{equation}
T_{\theta_1\to \theta_2}=T(\theta_2)-T(\theta_1)
\end{equation}
 where $T(\theta_i)$ is defined through
\begin{align}
	T(\theta_i):=&-\dfrac{E}{2u\ell}\left[ \dfrac{(e-1)\ell+E\ln(1-\cos \theta_i)}{(e-1)^2}+\dfrac{(e\ell+\ell -E)\ln(1-\cos \theta_i)}{(e+1)^2}\right.\cr 
	&\left. +\dfrac{2e}{(e^2-1)^2}\left(\dfrac{E(e^2-1)}{-1+e\cos \theta_i}+(\ell -e^2\ell -2E)\ln(1-\cos \theta_i)\right)\right].  
\end{align}
Here $E$ and $e$ are defined through (\ref{defEe}).

The case of `source current' can also be studied when $w\neq u$. The results are qualitatively the same as we have obtained previously, therefore, we will not study them in detail. Instead, we consider another type of current: `the sink current'. 

\section{Sink}
\begin{figure}[htpb]
	\centering
	\includegraphics[width=0.75\linewidth]{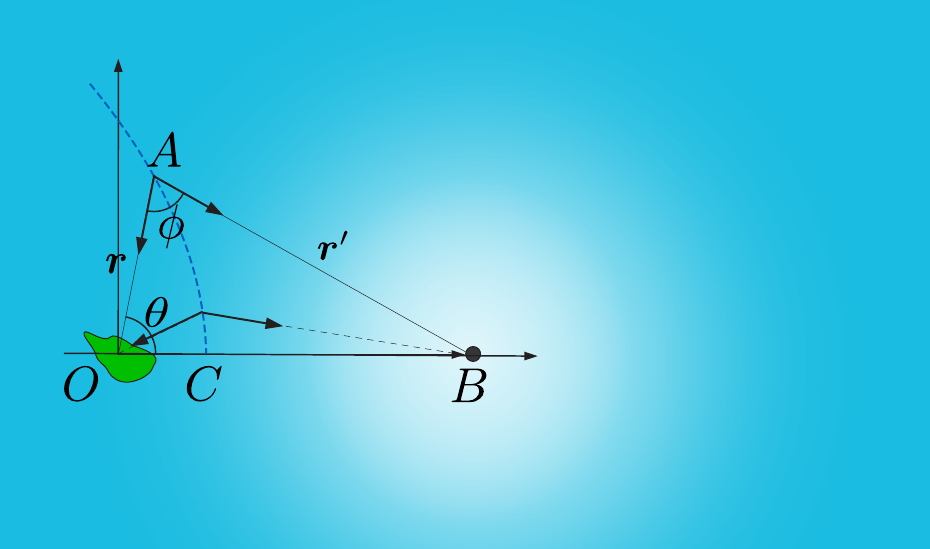}
	\caption{   The swimmer's path 	is a hyperbola with the sink and destination ($B$ and $O$) as its foci, if he/she swims toward destination.	 		}
	\label{f9}
\end{figure}

Let us assume there are a sink on the bottom of the lake that cause the water on the lake surface to move toward the point $B$ with a radial velocity of $u$. Let us assume that $w$ is equal to $u$. This is very similar the preceding one. The depth of the lake around the sink should vary  proportional to $r^{-1}$, where $r$ is the distance to the sink. 
The velocity of the swimmer has two parts, $-u \dfrac{\bm{r}}{|\bm{r}|}=-u {\bm e}_r$ 
and $-u \dfrac{\bm{r}'}{|\bm{r}'|}=u {\bm e}_{r'}$, where ${\bm e}_r$ ( ${\bm e}_{r'}$)
is the radial unit vector with respect to the point $O$ ($B$).  See figure (\ref{f9}).
The velocity components of the swimmer along $OA$ and $BA$ are
\begin{align}
	&\dot r= -u(1+\cos \phi),\\
	&\dot r'= -u(1+\cos \phi),
\end{align} 
which gives $\dot r-\dot r'=0$. So the swimmer moves on a path for which the difference of the its distance from two points $B$ and $O$ is constant. This is the characteristic of the hyperbola: $r-r'$ will be a constant. The swimmer's path is a hyperbola with $O$ and $B$ as its foci. Then path can be obtained to be 
\begin{equation}
	r=\frac{\ell^2-L^2}{2(L+\ell\cos \theta)}=:\frac{E}{1+e\cos \theta},
\end{equation} 
where
\begin{equation}
	E:= \frac{\ell^2-L^2}{2L}, \quad e:=\frac{\ell}{L}.
\end{equation} 
As the swimmer approaches the $x$ axis, at point $C$, its velocity tends to zero, where it eventually stops.

\section{Strategy to arrive at the island when source (sink) is present}
\begin{figure}[htpb]
	\centering
	\includegraphics[width=0.55\linewidth]{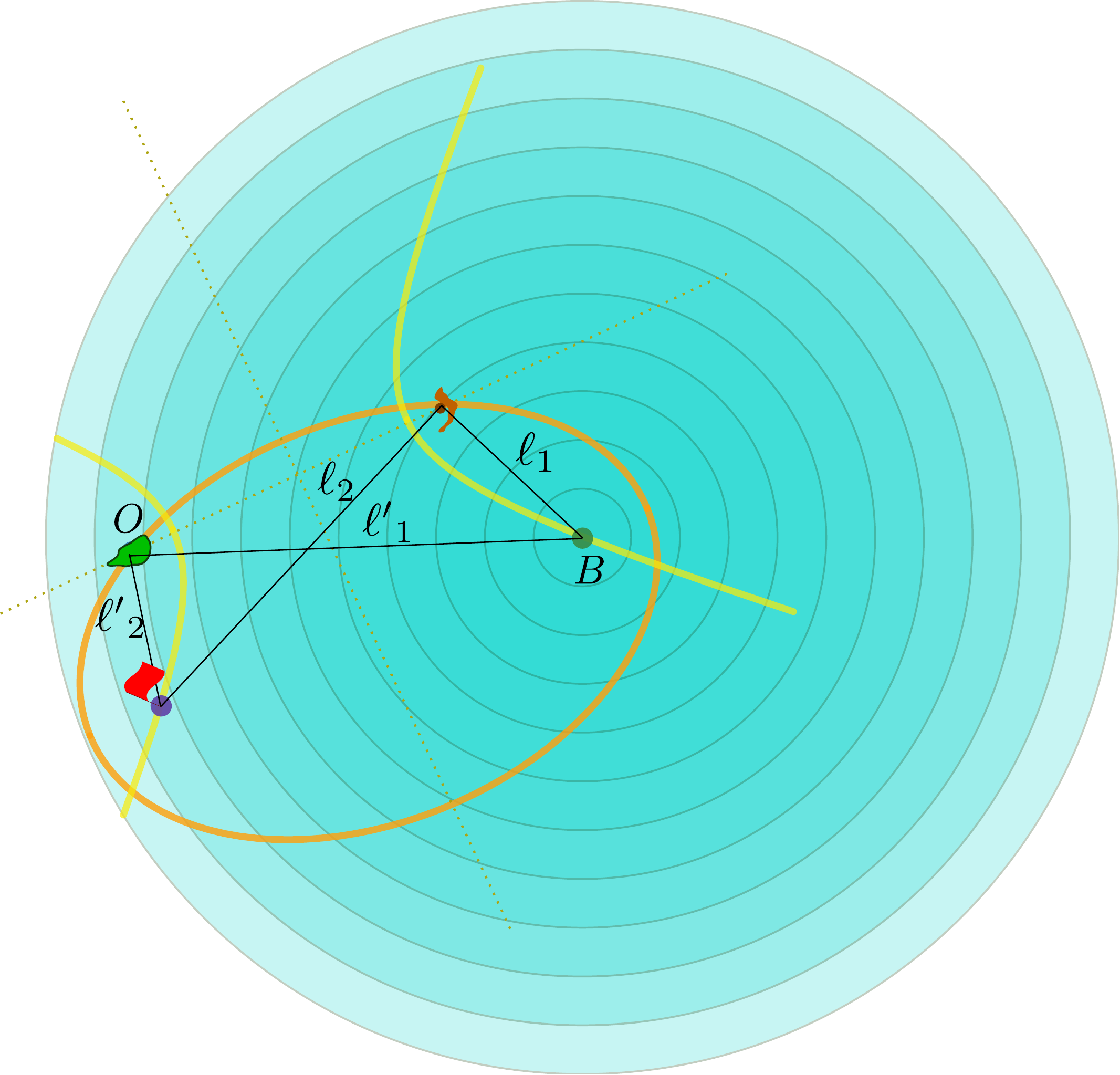}
	\caption{   Swimming in the presence of a source using a flag as a guide. The path will be an ellipse with the source and flag as its foci.  The loci of the points we should set the flag is  a hyperbola (shown by yellow line) for which two points, initial and destination are its foci.
	}
	\label{f10}
\end{figure}

In previous sections, we showed that when the swimmer's speed equals the drift, aiming at the island does not result in arrival at it, rather, the swimmer follows a part of an ellipse or a parabola and never reaches it. Despite being driven away (towards) a sink (source), a swimmer can manage to reach the island. Therefore, again a new strategy would be needed to follow. We take the new strategy as state in the case of uniform flow:  swimming towards a third point, the guide,  whose coordination is to be determined. This means that we put a fictitious flag somewhere and tell the swimmer to swim towards this flag. Let us concentrate on the case that a source is present at the stream. With similar reasoning presented in the previous sections, we can argue that the path of the swimmer is a part of an ellipse whose focal points are the source point and the fictitious flag. As we are interested in reaching the island, we have to choose a proper coordinate for the flag so that the emerging ellipse passes the island. See figure (\ref{f10}). 
We call the distances of the initial coordinate of the swimmer to the source and the flag by $\ell_1$ and $\ell_2$, and the distances of the island to the same points by ${\ell'}_1$ and ${\ell'}_2$. As the path is a part of an ellipse, we have 
$\ell_1+\ell_2={\ell'}_1+{\ell'}_2$ which means 
$\ell_2-{\ell'}_2={\ell'}_1-\ell_1$. The right handside of this equation is a fixed and known quantity, therefore, one concludes that we have to put the flag at a point whose distance from the swimmer's initial coordinates and the island is this fixed quality. This makes a hyperbola whose foci are these two points. Note that in the special case where the flag is put at infinity, the path of the swimmer is not an ellipse, rather, it is a hyperbola.

For the case of a sink, with very similar steps one can show that the flag should be put at an arbitrary point on an ellipse whose foci are the swimmer's initial coordinates and the island. The interesting question is which choice leads to the fastest arrival to the island.  
\subsection{The optimum path}
\begin{figure}[htpb]
	\centering
	\includegraphics[width=0.6\linewidth]{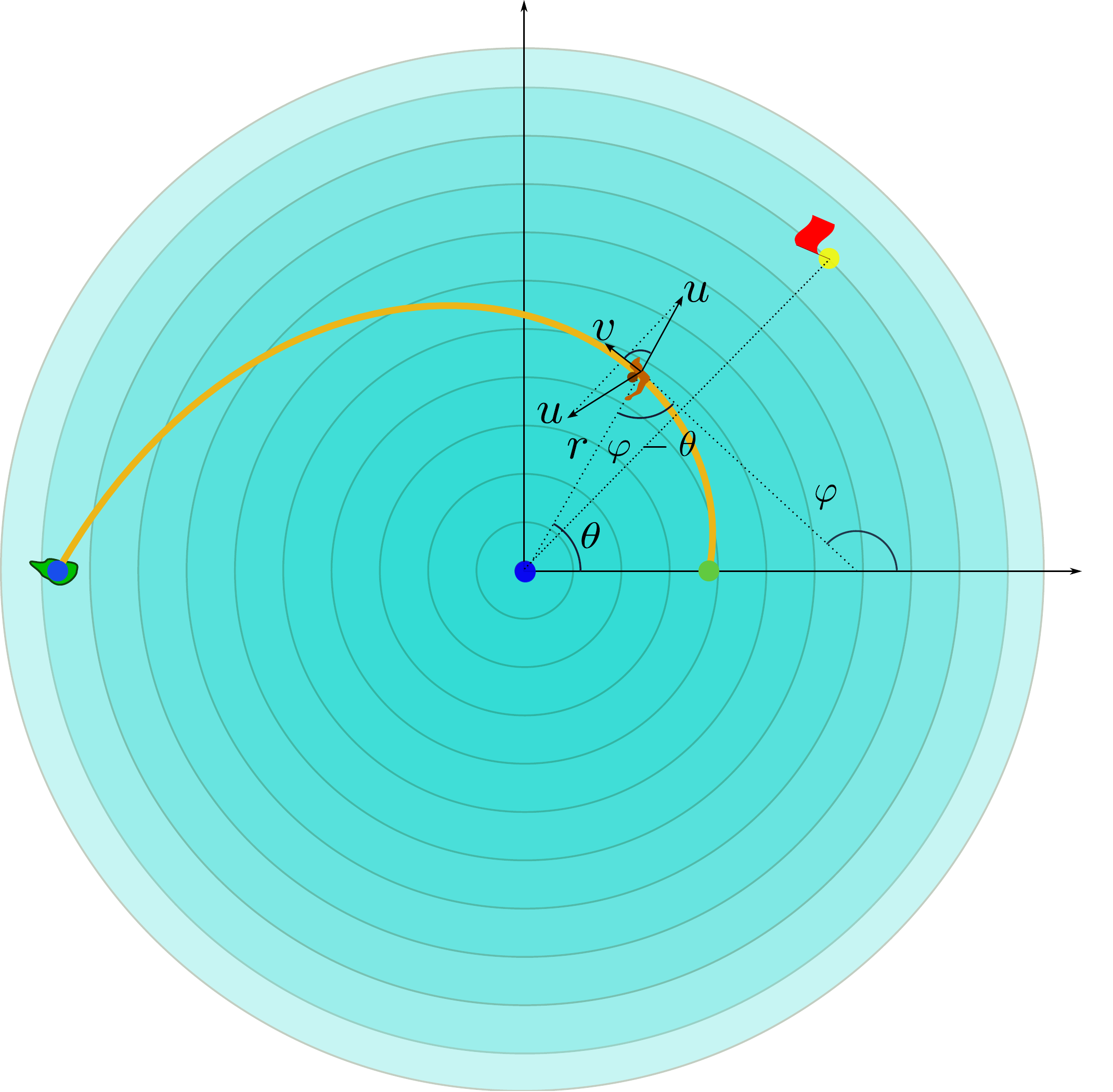}
	\caption{    Swimming toward a destination in the presence of a source. The path of least time is shown by orange line.  		
	}
	\label{f11}
\end{figure}

In this step, we will investigate path of least time. Let us take the spring at the origin. The path  with the shortest time is shown by the orange line in figure (\ref{f11}). It is more easier to use polar coordinates. 
 Denoting the swimmer's  velocity by $v=2u\cos(\varphi -\theta)$, and the differential path by ${\rm d}s=\sqrt{	({\rm d}r)^2+	(r{\rm d}\theta)^2}$, the swimming time is
\begin{align}
	T=&\int {\rm d}t=\int \frac{{\rm d}s}{v}=\int \frac{{\rm d}s}{2u\cos(\varphi -\theta) }\cr\label{var2}
	=&\dfrac{1}{2u}\int \frac{{\rm d}s}{{\rm d}r/{\rm d}s}=\dfrac{1}{2u}\int \frac{(r^2+r'^2)\, {\rm d}\theta}{r'}\cr
	=:& \dfrac{1}{2u}\int J(r,r')\, {\rm d}\theta.
\end{align}
Here the prime means differentiation with respect to $\theta$. Note that 
\begin{equation}
	{\rm d}r={\rm d}s\,\cos(\varphi -\theta)
\end{equation}
As $J(r,r')=\dfrac{r^2}{r'}+r'$ does not explicitly depend on $\theta$,
we arrive at 
\begin{align}
J-r'	\dfrac{\partial J}{\partial r'} ={\rm constant}.
\end{align}
This gives 
\begin{align}
	r'=Cr^2,\qquad \Rightarrow \qquad r=\dfrac{1}{A-C\theta}.
\end{align} 
Here $A$ and $C$ are two constants can be obtained using the arrival and destination coordinates. To interpret the path one may use the well-know method of image. The yellow point with the red flag is the image of the source with respect to the tangent line to the path at the point of swimmer. The swimmer should move away from the flag. As the swimmer displaces on the path the image will move too.
If a sink has replaced the source at the origin, principally, the path of least time does not change. Although it needs to change ${\bm v}$  by $-{\bm v}$, and replace the start point and the destination. 

\section{More sources and sinks}

Up to now, we have considered the problem of a swimmer in the presence of just one source/sink. As a mathematical problem, one can generalize the situation to the case where more sinks and sources are present in the lake. Note that even the parallel current case, can be obtained by putting a source/sink at infinity. Furthermore, the velocity of the swimmer  itself (when the swimmer takes the strategy of swimming toward the destination) can also be interpreted as a sink at the island while the swimmer just follows the total stream velocity at each point. 

\begin{figure}[htpb]
	\centering
	\includegraphics[width=0.322\linewidth]{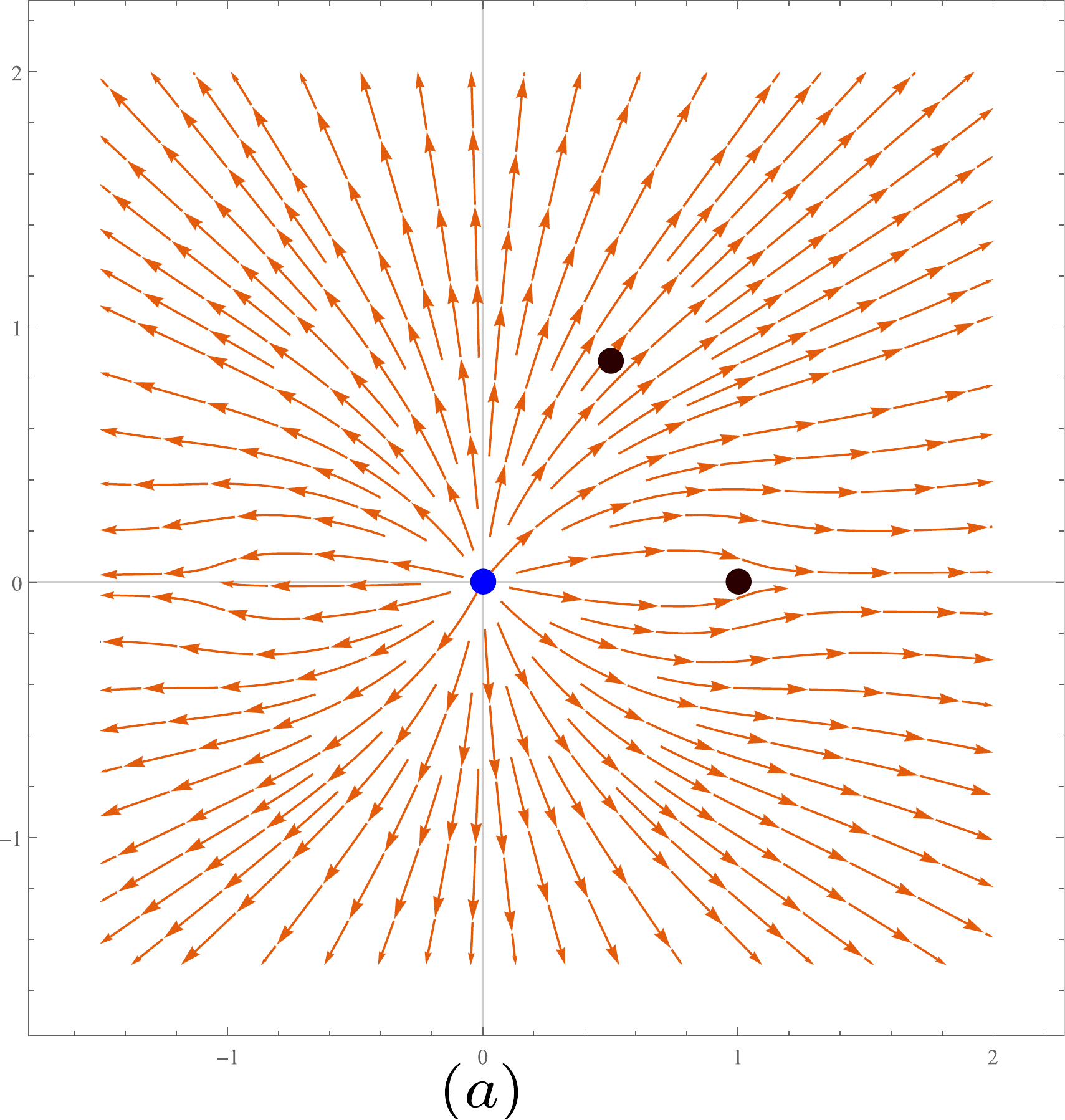}
	\includegraphics[width=0.32\linewidth]{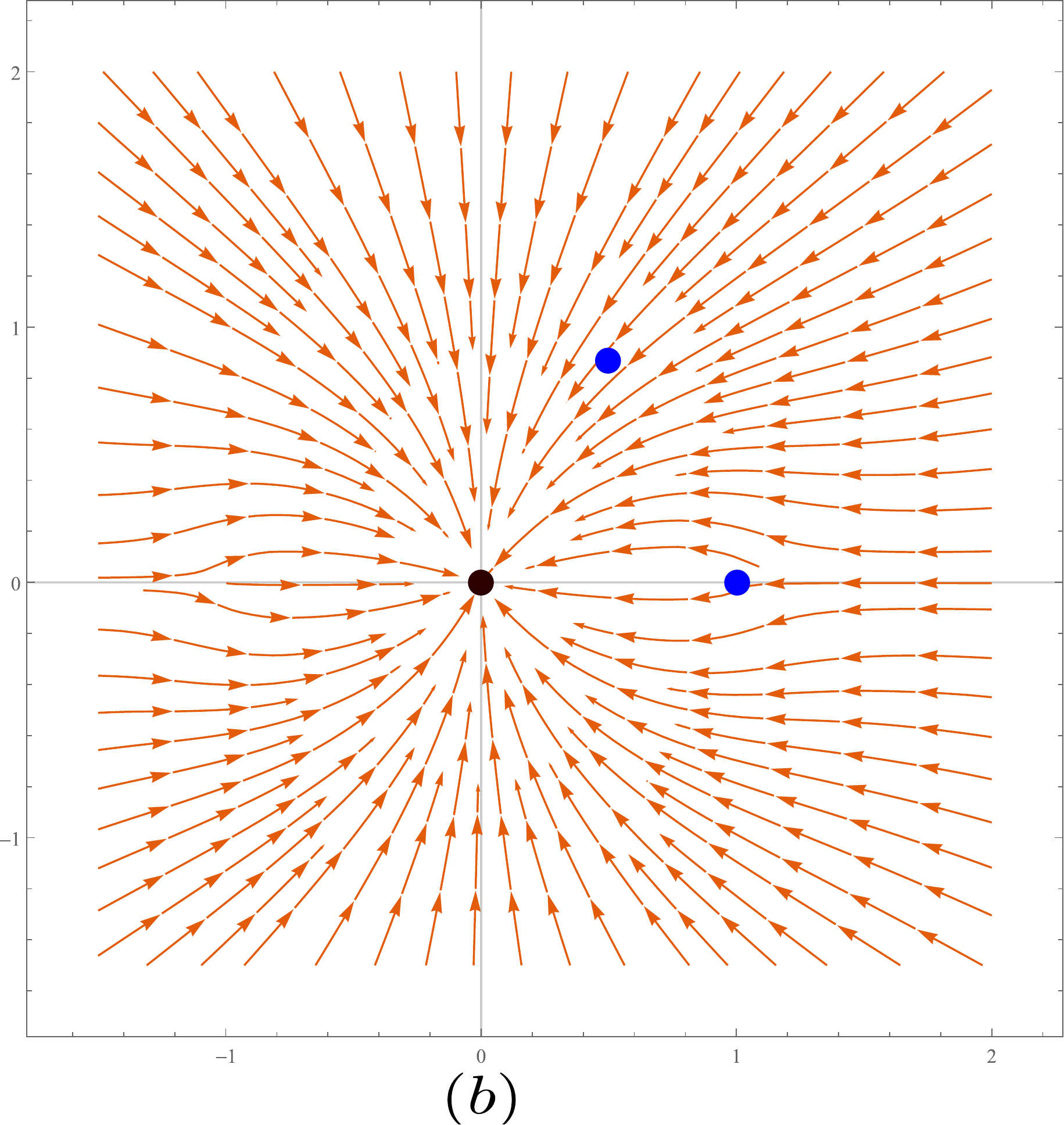}
	\includegraphics[width=0.323\linewidth]{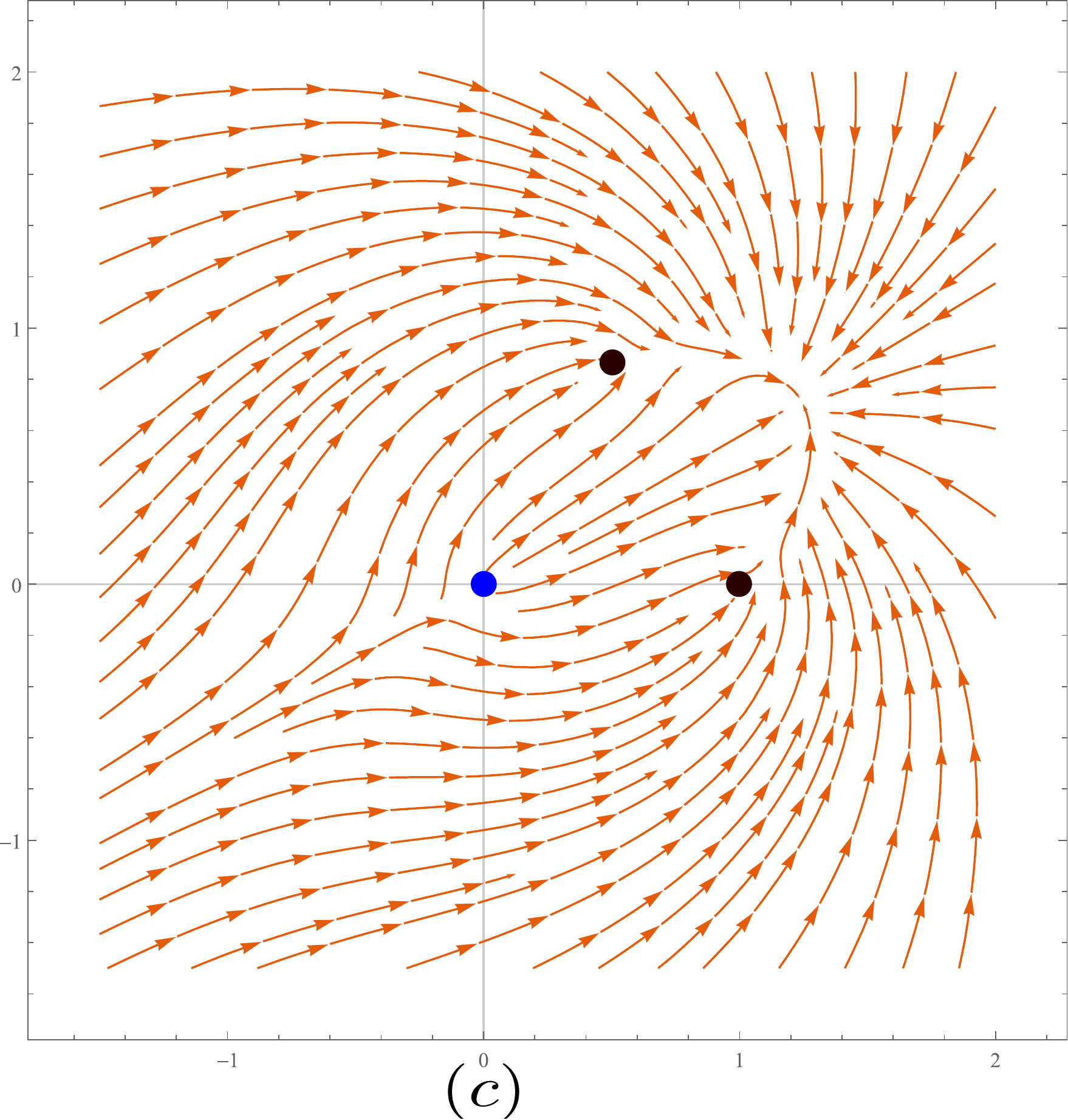}
	\caption{   The phase portraits of a system of three sources/sinks. The blue(black) dots represent sources(sinks). 
In case $(a)$, the positive velocity of the source is greater than the magnitude of the sum of the sink velocities. Therefore, no matter where the swimmer begins,  he/she will finally flow with the stream towards infinity. In case $(b)$, a sink is more powerful than the sum of other sinks/sources, the swimmer's path ends in the sink. In case $(c)$, none of the sources or sinks dominates the sum of the other two. 		
	}
	\label{three-phase}
\end{figure}

 Therefore, to go beyond what we have studied previously,  the total number of sink and source is at least three. We have to remember that at least have one sink should be present in the system (due to the velocity of the swimmer). The simplest cases are those which have exactly three total sources/sinks. We will call them the source/sink A$_1$, A$_2$, and A$_3$. Each of them produces a stream away from/towards the corresponding point with the velocity $v_i$ with $i\in \{1,2,3\}$. We may consider these velocities to be either positive or negative to show if the point is a source or a sink. Moreover, we may sometimes refer to these velocities as the strength of the sink/source. Three typical cases are shown in figure (\ref{three-phase}). Generally, in such cases, the resulting path of the swimmer cannot be obtained analytically. However, the system can be analyzed as a dynamical-system problem and one can ask: `what is the destiny of the swimmer?'.

Generally, the different destinies of the swimmer can be categorized in the following way: 
\begin{itemize}
	\item 1) It will flow towards infinity. 
	\item 2) It will arrive at one of the sinks.
	\item 3) The trajectory of the swimmer ends up somewhere in the lake where there is neither a sink nor a source.
\end{itemize}
Plotting the phase portrait of such systems helps to understand these different situations. Figure \ref{three-phase} shows the phase portraits of systems each belonging to one of the above categories. The coordinates of the three sources/sinks are considered to be on an equilateral triangle, while the velocities are taken to be $(1,-0.3,-0.3)$ and  $(-3,1,1)$ and $(1.5,-1,-1)$ in sub-graphs $(a)$, $(b)$, and $(c)$, respectively. In case $(a)$, the positive velocity of the source is greater than the magnitude of the sum of the sink velocities. Therefore, no matter where the swimmer begins,  he/she will finally flow with the stream towards infinity. On contrary, as in case $(b)$, when a sink is more powerful than the sum of other sinks/sources, the swimmer's path ends in the sink.
The more interesting case is when none of the sources or sinks dominates the sum of the other two. An example of such cases is sketched in figure (\ref{three-phase}c). The swimmer finally arrives at a point which is neither any of the sources/sinks nor infinity: a fixed point is formed in this case. Of course, in general, this fixed point needs not to be stable. For example, if in the same case as above, the velocity $v_1$ is chosen to be less than unity, the fixed point turns out to be an unstable one and the swimmer will eventually arrive at either the point b or c, depending on the initial condition. 

It is possible to find the necessary condition for the stability of infinity or a sink. Although we have considered the case of three sinks/sources, the same argument can be applied to cases with more sinks/sources.

{\bf Stability of infinity:} At infinity, all the sinks/sources can be considered to be at the origin in the first approximation. Therefore, the flow at infinity is a radial one with velocity $v_\infty=\sum_i v_i$. Hence, depending on whether $v_\infty$ is positive or negative, infinity would be stable or unstable.

{\bf Stability of a sink:} It is clear that a source cannot be stable. However, if a sink is strong enough, it can be a stable point. The velocity field in the vicinity of the sink $j$ can be written in the following form:
\begin{equation}
    v=v_j {\bm e}_r+\sum_{i\neq j} v_i {\bm e}_{ji}= v_j {\bm e}_r+ v_{\rm other}{\bm s}.
\end{equation}
where, $ {\bm e}_{ji}$ is the unit vector in the direction ${\bm r}_j-{\bm r}_i$. To be specific, the term $v_{\rm other}{\bm s}$ is velocity field due to other points at the point $j$, we have chosen ${\bm s}$ in a way that $v_{\rm other}> 0$. If $v_{\rm other}<|v_j|$, the sink $j$ would be stable because the radial inward velocity dominates the flow due to the other points. However, if $v_{\rm other}>|v_j|$ the sink is not a fixed point at all. 

\begin{figure}[h]
	\centering
	\includegraphics[width=0.6\linewidth]{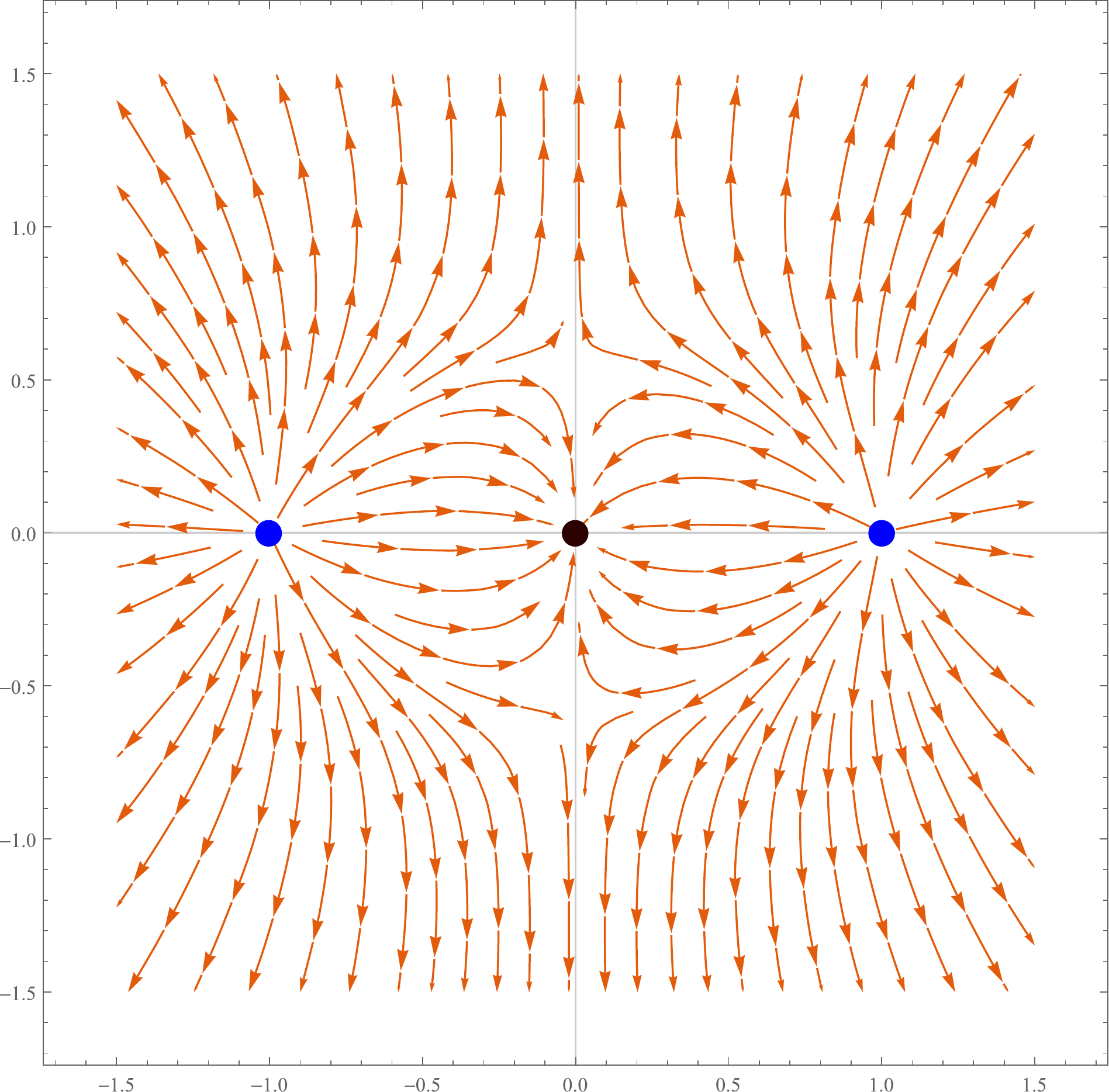}
	\caption{   A system of two sources with strength $v=1$ on points $(-1,0)$ and $(1,0)$ and a sink with strength $v'=-1$ on the origin. The system has both infinity and origin as stable fixed points. There are also two saddle fixed points on the $y$-axis.	
	}
	\label{two-fixed}
\end{figure}

The third case, where the fixed point is not at infinity or on a sink cannot be treated so easily. One has to find the fixed point first, and through the derivative matrix find out if the point is stable or not or if it is a saddle point. It is worth mentioning that it is possible to have several fixed points in a system of different types. For example, consider the case where two sources with strength $v=1$ on points $(-1,0)$ and $(1,0)$ and a sink with strength $v'=-1$ on the origin. Both infinity and origin are stable fixed points and there are two saddle fixed points on the $y$ axis. See figure (\ref{two-fixed}).

\section{Concluding remarks}
We have considered the path of a swimmer in the presence of different background surface currents. The problem shows both conceptual and mathematical elegant points. One has to exploit some geometrical and also algebraic tools to acquire a good understanding of the system. Additionally, the concept of taking non-trivial strategy to arrive at the desired point in a given background current make it quite interesting and challenging to an undergraduate student. Moreover, analyzing the system using dynamical system tools help the student see the whole system from a completely different point of view. 
\\[\baselineskip]
\textbf{Acknowledgment}: The work of A. Aghamohammadi was supported by the
research council of the Alzahra University. A. A. would like to thank M. Khorrami for useful comments.

\end{document}